\documentclass[10pt]{article}

\usepackage[utf8]{inputenc}
\usepackage{enumitem, graphicx}
\usepackage{authblk}
\usepackage{chronology}
\usepackage[margin=1in]{geometry}
\usepackage{blindtext}
\usepackage{hyperref}
\usepackage{indentfirst}
\usepackage[utf8]{inputenc}
\usepackage{subfig}
\usepackage{float}

\title{\textbf{Comprehensive Forecasting of California's Energy Consumption: A Multi-Source and Sectoral Analysis Using ARIMA and ARIMAX Models}}

\author[3]{Zahra Moslemi}
\author[1]{Logan Clark}
\author[2]{Sarah Kernal}
\author[1]{Samantha Rehome}
\author[1]{Scott Sprengel}
\author[3]{Ahoora Tamizifar}
\author[4]{Shawna Tuli}
\author[4]{Vish Chokshi}
\author[4]{Mo Nomeli}
\author[4]{Ella Liang}
\author[4]{Moury Bidgoli}
\author[4]{Jeff Lu}
\author[4]{Manish Dasaur}
\author[4]{Marty Hodgett}

\affil[1]{\textit{California State University, Fullerton}}
\affil[2]{\textit{Cypress College}}
\affil[3]{\textit{University of California, Irvine}}
\affil[4]{\textit{Accenture California}}

\date{August 2023}

\usepackage[backend=biber]{biblatex}
\begin{document}
\maketitle
\begin{center}
    \rule{6.5in}{1.5pt}
\end{center}
\medskip

\begin{abstract}
California's significant role as the second-largest consumer of energy in the United States underscores the importance of accurate energy consumption predictions. With a thriving industrial sector, a burgeoning population, and ambitious environmental goals, the state's energy landscape is dynamic and complex. This paper presents a comprehensive analysis of California's energy consumption trends and provides detailed forecasting models for different energy sources and sectors \cite{international2009world}. The study leverages ARIMA \cite{shumway2017arima} and ARIMAX models \cite{peter2012arima}, considering both historical consumption data and exogenous variables. We address the unique challenges posed by the COVID-19 pandemic and the limited data for 2022, highlighting the resilience of these models in the face of uncertainty. Our analysis reveals that while fossil fuels continue to dominate California's energy landscape, renewable energy sources, particularly solar and biomass, are experiencing substantial growth. Hydroelectric power, while sensitive to precipitation, remains a significant contributor to renewable energy consumption. Furthermore, we anticipate ongoing efforts to reduce fossil fuel consumption. The forecasts for energy consumption by sector suggest continued growth in the commercial and residential sectors, reflecting California's expanding economy and population. In contrast, the industrial sector is expected to experience more moderate changes, while the transportation sector remains the largest energy consumer.
\end{abstract}

\section{Introduction}
California is the second largest consumer of energy in the Unites States\cite{outlook2010energy}. This is due to rapid urbanization and a thriving industrial sector. California is also the most populous state in the nation and has the largest economy in the United States. Thus, accurate predictions of the state's energy demands are crucial in order for policymakers to make decisions to help shape a sustainable future \cite{outlook2010energy}.\medskip

California has long been a trailblazer in the shift toward cleaner and more sustainable energy sources \cite{kirkpatrick2014promoting}. The state ranks as the second-highest producer of electricity from renewable energy in the United States, underscoring its commitment to a greener future. Recent policy initiatives further emphasize this transition, promoting the use of renewable energy while imposing limitations on fossil fuel consumption \cite{ghanadan2005using}. California's ambitious targets include becoming 60\% reliant on renewable energy by 2030 and striving for 100\% clean energy by 2045, a testament to its dedication to environmental sustainability.

In spite of the growing popularity of renewable energy, fossil fuels continue to dominate California's energy landscape. The state stands as the second-largest consumer of natural gas and petroleum in the nation \cite{hoots1935natural}, contributing to approximately 8\% of the total consumption of refined petroleum products in the United States \cite{outlook2010energy}. This persistent reliance on fossil fuels can be attributed, in part, to California's extensive transportation infrastructure. With more registered motor vehicles and a higher volume of vehicle miles traveled by residents than any other state, the demand for traditional fuels remains substantial.

\subsection{Dataset Description}
\indent Our data come from various public institutions, including the \href{https://www.eia.gov}{\textit{United States Energy Information Administration(EIA)}}, \href{https://dof.ca.gov}{\textit{State of California Department of Finance(DOF)}}, and \href{https://www.ncei.noaa.gov/access/monitoring/climate-at-a-glance/statewide/time-series}{\textit{National Centers for Environmental Information: National Oceanic and Atmospheric Administration (NOAA)}}. This study is based on annual data, with values running from 1960 to 2021 for some variables and 1970 to 2021 for others.
\medskip
\newline \indent Total yearly energy consumption for California is given in trillion British thermal units (Btu), where one Btu is “the quantity of heat required to raise the temperature of one pound of liquid water by 1° Fahrenheit (F) at the temperature that water has its greatest density (approximately 39° F)” \cite{outlook2010energy}.In this dataset, yearly energy consumption is also broken down by energy sources (coal, natural gas, petroleum, biomass, geothermal, solar\cite{denholm2015overgeneration, hamzehconceptual}, wind \cite{dvorak2010california}, and hydroelectric) and sectors (commercial, industrial, residential, and transportation).

\medskip
Additional numerical variables include California's population; energy production for California, broken down by source (trillion Btu); average annual price of energy in California, broken down by source and sector (dollars per million Btu); annual precipitation in California (inches), and average maximum temperature in California (degrees Fahrenheit).

\medskip
\indent The one categorical variable utilized in this study is that providing the Mnemonic Series Name (MSN) for a recorded observation of energy consumption.
Each MSN is created in terms of several classifiers, including energy source, the sector or activity to which the corresponding observation of consumption is dedicated, and the units this value is to be expressed in. The MSN variable allows us to break down energy consumption in California by these classifiers.

\subsection{Goals}
Our project seeks to predict energy consumption in California in the following way:

\begin{itemize}
    \item Predict fossil fuels and renewable energy consumption for the next 10 years.
    \item Predict energy consumption by sector for the next 10 years.
\end{itemize}

\section{Materials and Methods}

\subsection{Exploratory Data Analysis}
We began our exploratory data analysis by looking into California's fossil fuel consumption over time, taking note of the largest sources of consumables as well as significant peaks in our exploratory plots. Upon first glance, it was clear natural gas and petroleum were California's primary sources of fossil fuel consumption. Before going further, we decided to take a step back and look at historic events that have influenced oil in California that could explain the volatility of our plots \cite{tennyson2005growth}. 

\begin{figure}
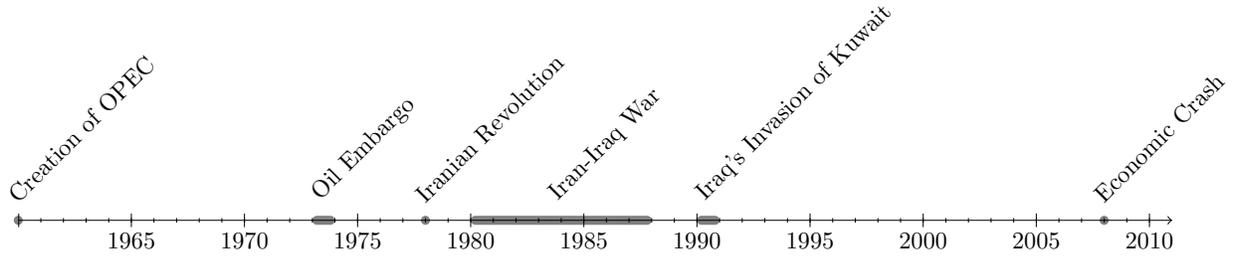

    \begin{chronology}[5]{1960}{2010}{\textwidth}
        \event{1960}{Creation of OPEC}
        \event[1973]{1974}{Oil Embargo}
        \event{1978}{Iranian Revolution}
        \event[1980]{1988}{Iran-Iraq War}
        \event[1990]{1991}{Iraq's Invasion of Kuwait}
        \event{2008}{Economic Crash}
    \end{chronology}
    \captionof{figure}{Timeline of Key Historical Events Impacting Fossil Fuel Consumption (1960-2010) in California and Beyond. Events include the creation of OPEC, the 1973 Oil Embargo, the Iranian Revolution in 1978, the Iran-Iraq War (1980-1988), Iraq's Invasion of Kuwait in 1990, and the Economic Crash of 2008. Gray lines represent significant events, while gray shaded regions indicate the duration of certain impactful events.}
\end{figure}

Our exploration of California's historical oil landscape \cite{leifer2006tracking, maugeri2006age} begins with the establishment of the Organization of the Petroleum Exporting Nations (OPEC) in 1960 \cite{desta2003organization}. OPEC's primary objective was to stabilize and defend global oil prices. Over a decade later, we have the oil embargo beginning in 1973, following President Nixon's request for Congress to make $2.2$ billion dollars in emergency aid to Israel for the Yom Kippur War. Because of this, the Organization of Arab Petroleum Exporting Countries initiated an oil embargo on the United States, limiting our imports of oil products. This lasted around two years.

The Iranian Revolution of 1978 and the subsequent Iran-Iraq war in the 1980s caused substantial disruptions to global oil production\cite{alnasrawi1994economy, long2012oil}. The Iranian Revolution of 1978 caused the oil output in Iran to drop over $5$ million barrels a day, leading to a $5$ percent global loss in production. Much like the Iranian Revolution, the Iran-Iraq war caused a sharp drop in oil production for most of the eighties. Iraq invaded Kuwait in 1990 causing the Bush administration to release $34$ million barrels of oil from the U.S Strategic Petroleum Reserve anticipating an oil crisis. To conclude the timeline, oil prices steadily began to increase from 2006 to a record high in 2008, leading into the economic crash of 2008. The crash caused oil prices to plummet, resulting in a sharp decrease in profits for oil and gas companies\cite{hamilton2013historical}.

\begin{figure}[H]%
    \centering
    \includegraphics[scale = 0.7]{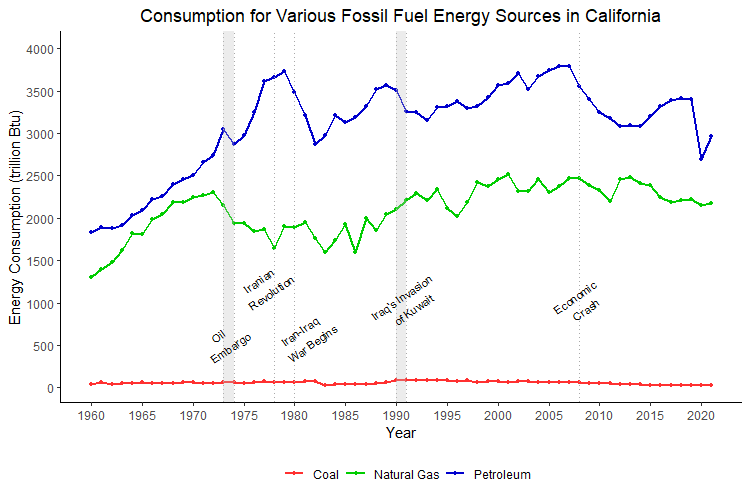}
    \caption{Fossil fuel energy consumption in trillion Btu from 1960 to 2021. Gray lines indicate events that had an impact on one or more fossil fuels in California or the rest of the US while gray shaded regions represent the duration of certain events\cite{ajanovic2020role}.}%
\end{figure}

We can see from the plot above that coal is consumed on a considerably smaller scale in California relative to natural gas and petroleum. When looking at a plot of California's coal production\cite{newcomb1978american}, it was nearly nonexistent, which is a major reason why the state doesn't consume much of it. In addition to coal, looking at the plot's timeline, a direct effect can be seen on the petroleum line following the major historic events of the Middle East. It is clear those events caused a momentary decline in the consumption of petroleum during those years.  

After reviewing the history of fossil fuels, we expanded our research to include the major policies that impacted the growth of renewable energy sources in California.

\begin{figure}
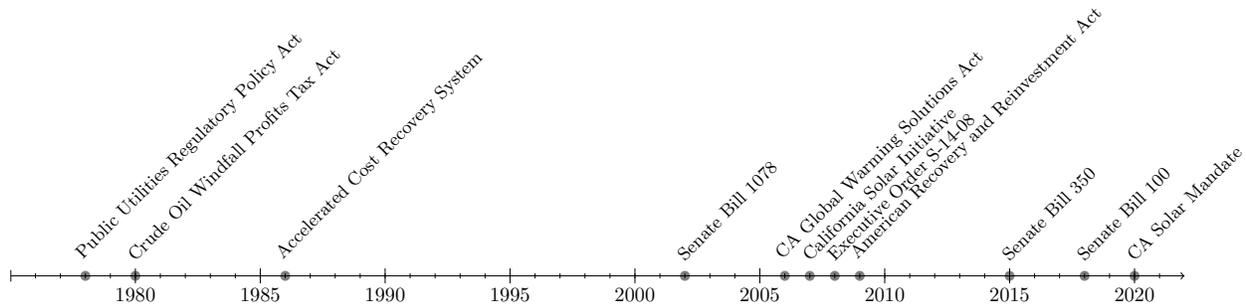

    \centering
    \begin{chronology}[5]{1975}{2021}{\textwidth}[20cm]
        \small
        \event{1978}{Public Utilities Regulatory Policy Act}
        \event{1980}{Crude Oil Windfall Profits Tax Act}
        \event{1986}{Accelerated Cost Recovery System}
        \event{2002}{Senate Bill 1078}
        \event{2006}{CA Global Warming Solutions Act}
        \event{2007}{California Solar Initiative}
        \event{2008}{Executive Order S-14-08}
        \event{2009}{American Recovery and Reinvestment Act}
        \event{2015}{Senate Bill 350}
        \event{2018}{Senate Bill 100}
        \event{2020}{CA Solar Mandate}
    \end{chronology}
    \caption{Chronology of Key Policies Shaping California's Transition to Renewable Energy (1975-2021)}
    \label{fig:chronology}
\end{figure}

The Public Utilities Regulatory Policy Act (PURPA) of 1978 played a significant role in opening doors for non-utility electricity generators, particularly renewable energy producers \cite{gonzalez2016evolving}. It allowed them to sell electricity into a utility's power grid, breaking the monopoly of power companies on electricity generation.

Following PURPA, the Crude Oil Windfall Profits Tax Act of 1980 was instrumental in promoting renewable energy. It authorized an increase in the nonresidential energy tax credit for solar and wind systems from 10\% to 15\% and raised the residential credit from around 22\% to 40\% . The Economic Recovery Tax Act of 1981 \cite{briner1981economic} established an Accelerated Cost Recovery System (ACRS) \cite{deloitte1981accelerated}, enabling wind, solar, and geothermal property placed in service after 1986 to be depreciated for tax purposes over a five-year period. These early acts laid the foundation for solar and wind energy to begin contributing to the power grid in the late 1980s.

In 2002, Senate Bill (SB) 1078 established California's renewable portfolio standard (RPS) program, aiming to increase the share of electric power sales from renewable energy sources. The program initially required 20\% of electricity retail sales to be served by renewables by 2017. California's commitment to environmental goals continued with the implementation of the Global Warming Solutions Act (GWSA) in 2006 \cite{visick2007if}. This act mandated a reduction in greenhouse gas emissions to 1990 levels by 2020 and the adoption of regulations to achieve that goal \cite{hanemann2007california}.

Further progress came with the California Solar Initiative (CSI) in 2007, which provided incentives for low-income customers to install solar photovoltaic systems \cite{hughes2015getting}. Executive Order S-14-08 increased RPS requirements, aiming for 33\% renewable energy by 2020 .

These early 2000s policies fostered investment in clean energy technology, making renewable energy generation more efficient and widespread. This laid the groundwork for a significant increase in renewable energy consumption, particularly evident in 2010. In 2015, Senate Bill 350 mandated a 50\% RPS by 2030, and in 2018, Senate Bill 100 increased the RPS to 60\% by 2030 and 100\% by 2045.

Effective from January 1, 2020, the California Solar Mandate is a building code requiring all new residential construction projects, including single-family homes, condominiums, and three-story apartment buildings, to have solar photovoltaic (PV) systems \cite{mani2010impact}. While the impact of the Solar Mandate may not yet be apparent in our data, it is likely to be a significant factor in the future growth of solar energy consumption \cite{10261977}.

Additionally, although not represented in the timeline, research by GTM (Greentech Media)\cite{hart2016deployment} indicated a remarkable reduction of 97.2\% in the average price of photovoltaic systems from 1975 to 2012, and the National Renewable Energy Laboratory's 2020 report\cite{karim2022national} revealed a drop of 64-82\% in solar system prices from 2010 to 2020. These price reductions, coupled with the array of clean energy policies, likely played a pivotal role in the exponential growth of solar energy consumption in California.

\begin{figure}[H]%
    \centering
    \includegraphics[scale = 0.7]{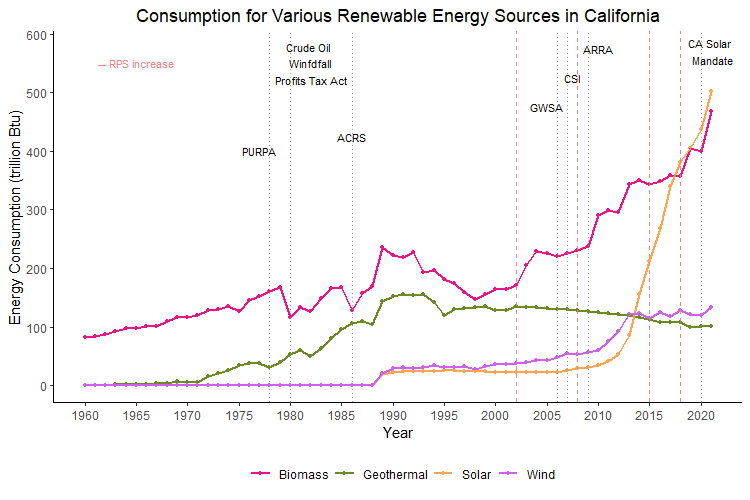}
    \caption{Renewable energy sources consumption in trillion Btu from 1960 to 2021. Gray lines indicate policies that had an impact on one or more renewable energy power sources in California or the rest of US while light red dashed lines represent an increase in the RPS.}%
\end{figure}

We can see from the plot above that until 2018 when solar energy surpassed it, biomass was the renewable energy source that California consumed the most. It is also apparent that until 1989 the amount of solar and wind energy consumed was negligible enough to be recorded as zero. Solar and wind energy both had a sudden increase around 2010, but while solar consumption continued to increase exponentially, wind power consumption leveled off after 2013. The plot also reveals that geothermal energy consumption has been slightly declining since about 1996.

Not included in the various renewable energy source consumption plot above was the line displaying hydroelectric power consumption. We did not include hydroelectric power because the volatility in hydroelectric power consumption from one year to the next created a very jagged line that distracted from the other renewable sources. The volatility in the consumption values did intrigue us though, and in our search for understanding we discovered that the pattern of peaks and lows in hydroelectric power consumption seemed to correspond to the peaks and lows of the average yearly precipitation in California.

\begin{figure}[H]
    \centering
    \includegraphics[scale = 0.6]{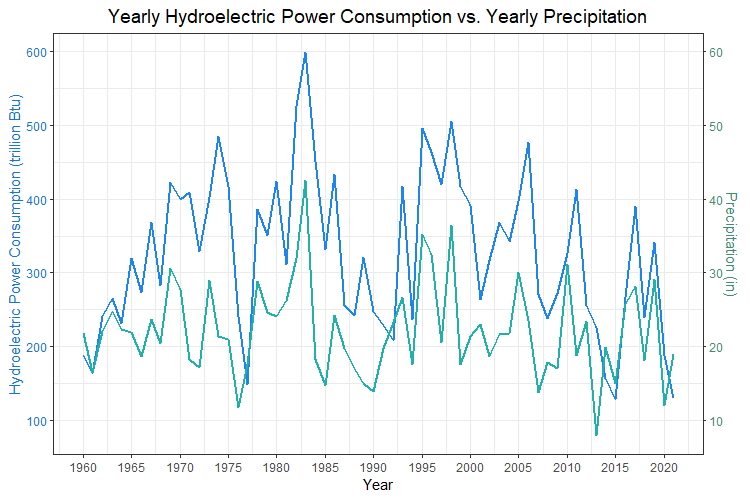}
    \caption{Hydroelectric power consumption in trillion Btu and average yearly precipitation in inches from 1960 to 2021. Blue line and left axis represents hydroelectric power consumption while teal line and right axis represents average yearly precipitation.}
\end{figure}

After researching the history of energy sources in California and plotting them to see their change over the years, we created a density plot with all the energy sources to see how much of California's total energy consumption comes from each source.

\begin{figure}[H]
    \centering
    \includegraphics[scale = 0.58]{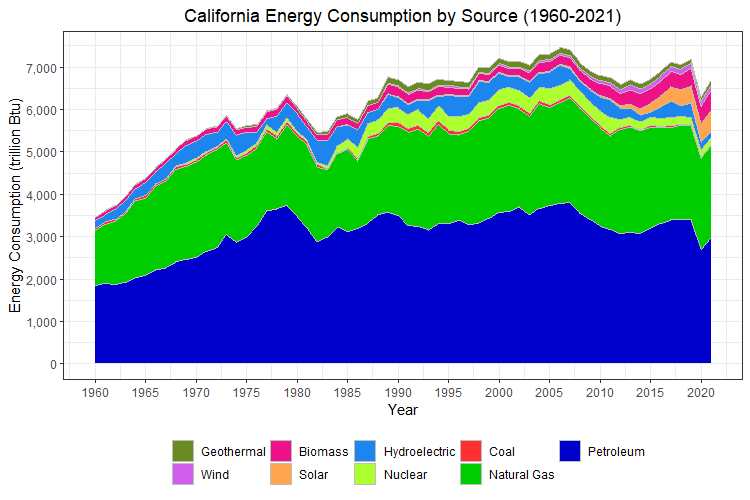}
    \caption{Portion of energy consumption in California in trillion Btu by source (geothermal, wind, biomass, solar, hydroelectric, nuclear, coal, natural gas, and petroleum) from 1960 to 2021.}
\end{figure}

The density plot clearly shows that energy from petroleum and natural gas account for about two-thirds of energy consumption in California while coal is barely a sliver of our consumption \cite{quigley1999decrease}. Nuclear energy accounted for a decent portion of California's consumption for a little over three decades, but it currently accounts for a very small portion. Renewable energy sources appear to only account for approximately 30\% of California's energy consumption. Historically, hydroelectric accounted for most of California's renewable energy consumption, but now it is only a very small portion. Solar is currently California's largest source of renewable energy consumption with biomass not far behind, but even combined, wind and geothermal energy account for very little of the total energy consumption.

We then created a density plot with all the energy sectors to see how much of California's total energy consumption comes from each sector.

\begin{figure}[H]
    \centering
    \includegraphics[scale = 0.58]{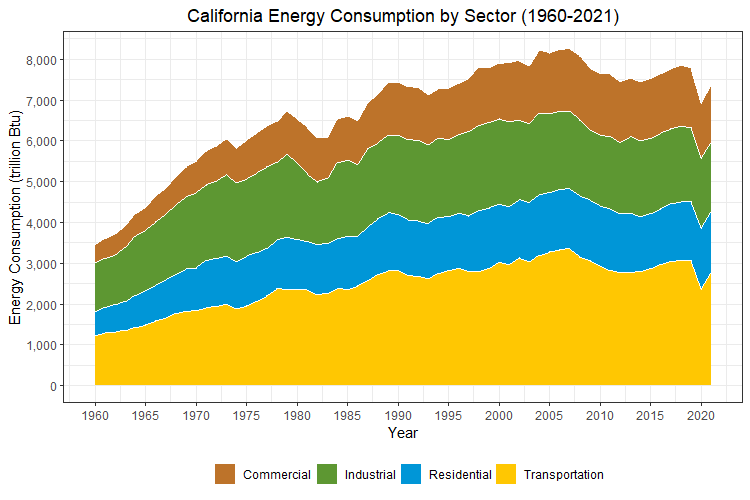}
    \caption{Portion of energy consumption in California in trillion Btu by sector (commercial, industrial, residential, and transportation) from 1960 to 2021.}
\end{figure}

The density plot clearly shows that California's transportation sector has historically been the largest energy consumer and currently accounts for a little over a third of California's consumption. The industrial sector is and has always been the next largest consumer and currently accounts for nearly a quarter of California's consumption. The commercial and residential sectors are the smallest energy consumers, each currently accounting for about a fifth of California's total energy consumption.

In this research, an R Markdown document authored by Logan Clark was employed to conduct a comprehensive analysis of energy consumption trends in California's industrial and transportation sectors. The document utilized various R packages, including tidyverse, janitor, dplyr, readr, tseries, and forecast, for data manipulation, analysis, and time series forecasting.

\subsection{Modeling and Implementation}
To properly model our data, we decided to use an ARIMA (Auto-regressive Integrated Moving Average) time series model. The ARIMA model takes in three separate parameters (p,d,q) \cite{reisen1994estimation}

\begin{itemize}
    \item p: The order of the AutoRegressive (AR) component, which represents the number of lag observations included in the model. The AR component is based on the idea that the past values in the time series can be used to predict future values.

    \item d: The order of differencing, which represents the number of times the raw observations are differenced to make the time series stationary. Stationarity is often required for time series analysis, and differencing is a common technique to achieve this.

    \item q: The order of the Moving Average (MA) component, which represents the number of lagged forecast errors in the prediction equation \cite{hamzehnew}. The MA component is based on the idea that the past forecast errors can help predict future values.
\end{itemize}
The equation given below is a general equation for all ARIMA time series models:

\medskip

\begin{equation} Y_t = \alpha + \beta_1Y_{t-1} + \beta_2Y_{t-2} + .. + \beta_pY_{t-p} + \epsilon_t + \phi_1\epsilon_{t-1} + \phi_2\epsilon_{t-2} + .. + \phi_q\epsilon_{t-q}
\end{equation}

\medskip

While this equation may seem complex and confusing, below is a quick break down of each of the variables and what they mean in the context of the equation

\medskip

\begin{itemize}
    \item $Y_t$ is the response variable at the current time period
    \item $\alpha$ is the constant or intercept term of the equation
    \item $\beta_p$ is the correlation of the value $Y_{t-p}$ has to the response variable. The values of the coefficient are between -1 and 1
    \item $\phi_p$ is the correlation of the value $\epsilon_{t-p}$ has to the response variable. The values of the coefficient are between -1 and 1
    \item $Y_{t-p}$ is the response variable $p$ periods ago
    \item $\epsilon_{t-q}$ is the error $q$ periods ago
\end{itemize}

Using the three parameters of the ARIMA model, we can make a forecasting model to predict future values of energy consumption based off of energy consumption values from previous years. We can also include other time series of the same time frame and frequency to account for other variables, such as California's population, energy production, energy prices, etc. Using the auto.arima() function in R, we're able to find the best parameters to fit our model by identifying the parameters with the lowest AICc values. We then test those models by excluding the last 5-10 years of consumption values to use as a testing set and training our model on the remaining values.

We tested the top few ($p$,$d$,$q$) parameters and included various exogenous predictors to see which would obtain the lowest mean-square error (MSE)\cite{prasad1990estimation} compared to the true testing data. We then chose the model parameters and variables with the lowest MSE to build an ARIMA model using the full dataset, which we then used to predict the next 10 years of values.

On the other hand, ARIMAX, or AutoRegressive Integrated Moving Average with eXogenous variables, is an extension of the ARIMA model that includes additional explanatory variables, also known as exogenous variables or covariates. The inclusion of these exogenous variables allows ARIMAX to capture the influence of external factors on the time series being analyzed.

The ARIMAX model is denoted as ARIMAX(p, d, q) with exogenous variables, where:

\begin{itemize}
    \item \textbf{p:} The order of the AutoRegressive (AR) component, representing the number of lag observations included in the model. AR relies on past values in the time series to predict future values.
    
    \item \textbf{d:} The order of differencing, indicating the number of times raw observations are differenced to achieve stationarity. Stationarity is crucial for time series analysis, and differencing is a common technique.
    
    \item \textbf{q:} The order of the Moving Average (MA) component, indicating the number of lagged forecast errors. MA incorporates past forecast errors to predict future values.
    
    \item \textbf{X:} Exogenous variables that are external to the time series but can influence it. These variables are not predicted by the model but are included to improve prediction accuracy.
\end{itemize}

\medskip

The general equation for ARIMAX time series models is as follows:

\medskip

\begin{equation}
Y_t = \alpha + \beta_1 Y_{t-1} + \beta_2 Y_{t-2} + \ldots + \beta_p Y_{t-p} + \epsilon_t + \phi_1 \epsilon_{t-1} + \phi_2 \epsilon_{t-2} + \ldots + \phi_q \epsilon_{t-q} + \gamma_1 X_{t-1} + \gamma_2 X_{t-2} + \ldots + \gamma_r X_{t-r}
\end{equation}

\medskip

Here's a breakdown of the variables in the ARIMAX equation:

\begin{itemize}
    \item $Y_t$: The response variable at the current time period.
    \item $\alpha$: The constant or intercept term of the equation.
    \item $\beta_p$: The correlation of the value $Y_{t-p}$ with the response variable, ranging between -1 and 1.
    \item $\phi_p$: The correlation of the value $\epsilon_{t-p}$ with the response variable, ranging between -1 and 1 \cite{rostami2023federated}.
    \item $Y_{t-p}$: The response variable $p$ periods ago.
    \item $\epsilon_{t-q}$: The error $q$ periods ago.
    \item $X_{t-r}$: Exogenous variable $r$ periods ago.
    \item $\gamma_r$: The coefficient representing the impact of the exogenous variable $X_{t-r}$ on the response variable.
\end{itemize}

The exogenous variables can represent external factors that are believed to have an impact on the time series. For example, in the context of forecasting sales, the ARIMAX model might include economic indicators, marketing expenses, or other relevant variables as exogenous inputs.

The parameters of the ARIMAX model, including the autoregressive parameters (\(\phi\)), moving average parameters (\(\theta\)), and coefficients associated with exogenous variables (\(\beta\)), are estimated using statistical methods, and the model performance is evaluated based on its ability to accurately predict future values of the time series.

\medskip

We also used Holt's damped linear trend method, a form of exponential smoothing, to forecast each renewable energy source separately. Exponential smoothing methods use weighted observations of past values to make forecasts. Holt's linear trend method is comprised of two smoothing equations, one for the level, and one for the trend. The dampening parameter is added to this method to help prevent over-casting future values by flattening out the curve as time goes on. This method was chosen for it's ease of use and ability to handle nonlinear and non-stationary data. 
\newline

Holt's Damped Trend Equation: 
\begin{equation}
\hat{y}_{t+h|t} = l_t+(\phi+\phi^2+...+\phi^h)b_t
\end{equation}

\begin{equation}
l_t = \alpha y_t+(1-\alpha)(l_{t-1} + \phi b_{t-1})
\end{equation}

\begin{equation}
b_t = \beta^*(l_t-l_{t-1})+(1-\beta^*)\phi b_{t-1}
\end{equation}
\newline
Equation (1) is the entire forecast equation of Holt's damped trend.  
\begin{itemize}
    \item $\hat{y}$ is the forecast value
    \item $l_t$ is the level estimate at time $t$
    \item $\phi$ is the dampening term
    \item $b_t$ is the trend estimate at time $t$
\end{itemize}
Equation (2) is the level equation.
\begin{itemize}
    \item $\alpha$ is the level smoothing parameter   
\end{itemize}
Equation (3) is the trend equation.
\begin{itemize}
    \item $\beta^*$ is the trend smoothing parameter
\end{itemize}

\section{Results}

In the pursuit of a thorough examination of energy consumption in California, a comprehensive strategy was employed involving a total of nine ARIMAX models. The predictive models comprised a (2,0,0) ARIMAX model specifically designed for total energy consumption forecasting. Additionally, two (2,1,0) ARIMAX models were crafted to analyze petroleum and natural gas consumption, accounting for scenarios with and without the impact of COVID-19 \cite{kahn2021impact, wang2022impact}. A (1,1,1) ARIMAX model was deployed to predict hydroelectric power consumption, while a (2,1,2) ARIMA model was applied to forecast consumption trends in Solar, Wind, Biomass, and Geothermal sources.

Furthermore, a sector-specific focus was incorporated into the analysis, with four distinct ARIMAX models tailored to different segments of the energy landscape. These included (4,1,0) ARIMAX models for both the commercial and residential sectors, as well as (2,1,2) ARIMAX models for the industrial and transportation sectors. This meticulous multi-model framework, encompassing nine ARIMAX models in total, was designed to provide a comprehensive and nuanced understanding of energy consumption dynamics across diverse sectors in California.

To begin our various resulting predictive models, we have included a model of total energy consumption in California with a 10 year consumption forecast based on a (2,0,0) ARIMAX model which used California's population, total primary energy production, gas prices (adjusted for inflation), and the average price of energy (adjusted for inflation) as exogenous variables.

\begin{figure}[H]
    \centering
    \includegraphics[scale = 0.55]{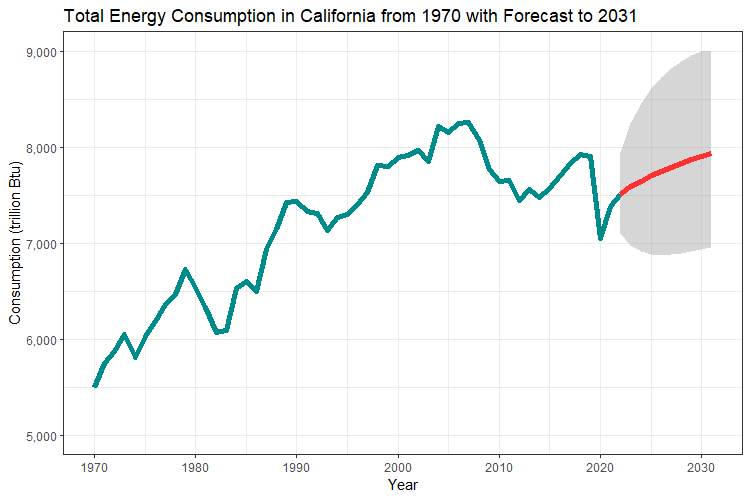}
    \caption{Total energy consumption in California in trillion Btu from 1970 to 2021. Dark cyan line represents the observed values, red line represents the forecast values, gray shaded area represents a 95\% confidence interval for the forecast values.}
\end{figure}

Based on the predictions of this model, California is expected to reach a consumption level of 7936.225 trillion Btu by 2031 with a 95\% confidence interval that spans from about 6952 to 9060 trillion Btu.

 Next, we look closer at each source, beginning with fossil fuels. Forecasting consumption of petroleum and natural gas utilized a (2,1,0) ARIMAX model that used population and production as exogenous predictors. Taking into consideration COVID's anomalous effect on energy consumption habits, we built another model on pre-COVID data and predicted the next 12 years.

\begin{figure}[H]%
    \centering
    \subfloat[\centering Including COVID Data]{{\includegraphics[width=7.5cm]{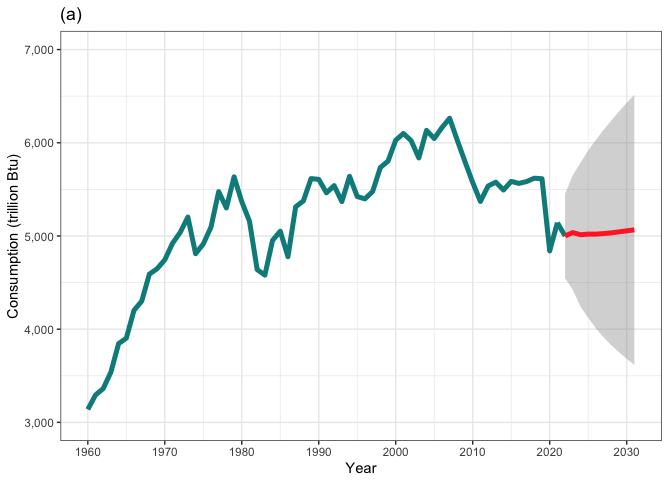} }}%
    \qquad
    \subfloat[\centering Excluding COVID Data]{{\includegraphics[width=7.5cm]{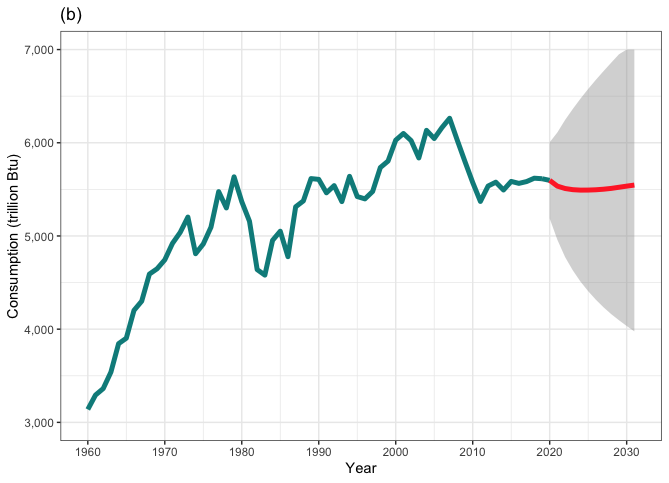} }}%
    \vspace{0.2cm}
    \caption{10 Year Fossil Fuel Consumption Forecasts. Blue line represents observed values, red line represents forecast values, gray shaded area represents a 95\% confidence interval for forecast values}%
\end{figure}

We predict fossil fuel usage will remain constant, but may decrease under new state policies. California's fossil fuel reduction efforts face challenges due to extensive transportation infrastructure and the ongoing demand for energy from various industries. Between the two models, we would expect consumption to fall closer in line to the one that excludes COVID data, as the extreme drop from COVID would skew predictions for future years, especially when using auto-regressive terms. 

We now move to renewable energy sources. Below are the resulting 10 year consumption forecasts of the (1,1,1) ARIMAX model that used precipitation as an exogenous predictor for hydroelectric power, and the (2,1,2) ARIMA model that was used to model the rest of the renewable energy sources. 

\begin{figure}[H]%
    \centering
    \subfloat[\centering Hydroelectric]{{\includegraphics[width=7.5cm]{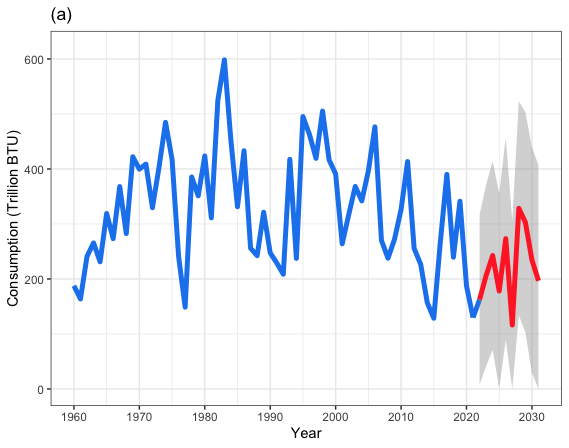} }}%
    \qquad
    \subfloat[\centering Solar, Wind, Biomass, and Geothermal]{{\includegraphics[width=7.5cm]{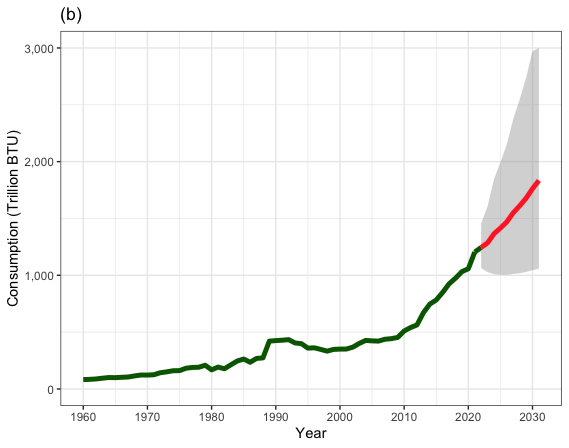} }}%
    \vspace{0.2cm}
    \caption{10 Year Renewable Energy Consumption Forecasts. Blue/Green line represents observed values, red line represents forecast values, gray shaded area represents a 95\% confidence interval for forecast values}%
\end{figure}

We anticipate that hydroelectric power will continue to be a major contributor to renewable energy consumption. However, it remains particularly sensitive to precipitation levels, making long-term predictions challenging. For hydroelectric power, more accurate forecasts are possible on a shorter timescale, ideally looking ahead just one year. By using short-term precipitation forecasts, we can provide a closer estimate for the upcoming year. Hydroelectric power has the potential to rise well above 350 trillion Btu during rainy seasons, but it could also drop below 200 trillion Btu during periods of drought.

Conversely, for other forms of renewable energy, we project a consistent increase in consumption as California strives to achieve its ambitious goal of 100\% renewable consumption by 2045. Among these renewable sources, we expect solar and biomass to be the primary contributors to this ongoing growth. This trend can be visualized in the subsequent plot, which employs Holt's dampened linear trend to offer a general forecast for the consumption of each distinct source in the near future. These forecasts were generated using a damped trend parameter of 0.95, and the entire process was automated through the use of the forecast package in R.

\begin{figure}[H]%
    \centering
    \subfloat[\centering Solar]{{\includegraphics[width=7.5cm]{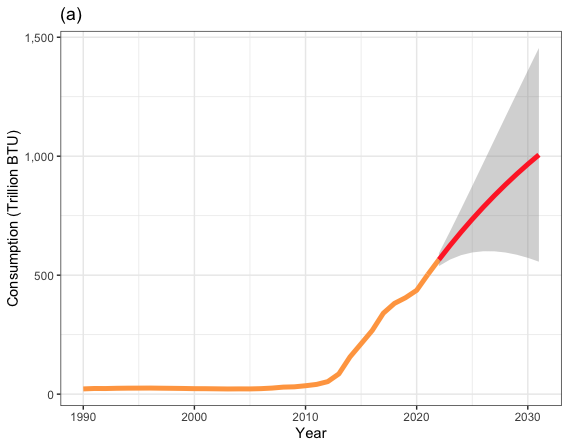} }}%
    \qquad
    \subfloat[\centering Wind]{{\includegraphics[width=7.5cm]{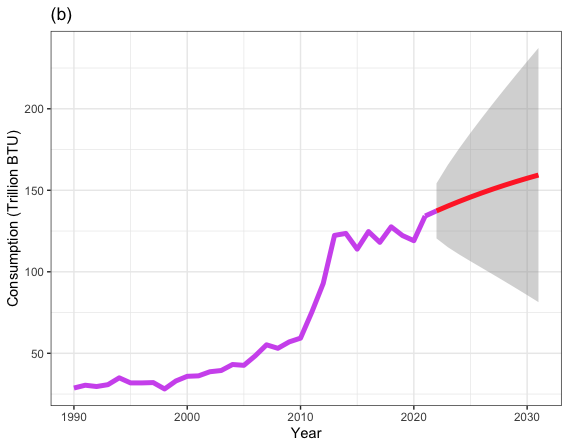} }}%
    \newline
    \centering
    \subfloat[\centering Biomass]{{\includegraphics[width=7.5cm]{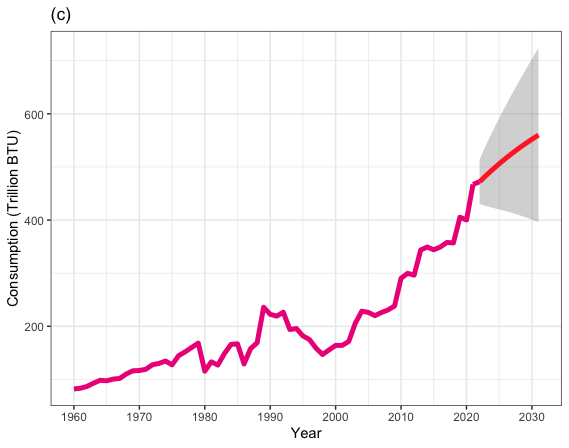} }}%
    \qquad
    \subfloat[\centering Geothermal]{{\includegraphics[width=7.5cm]{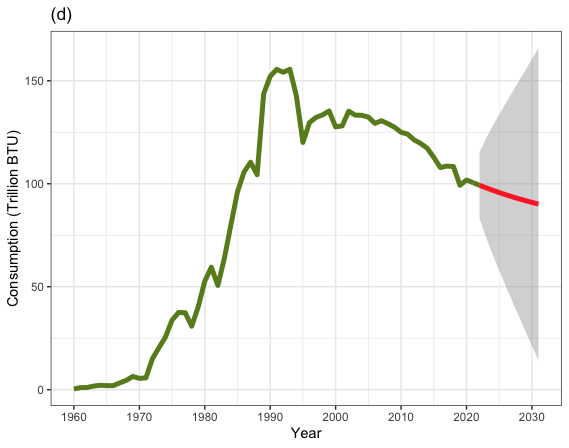} }}%
    \vspace{0.2cm}
    \caption{Red lines represent predicted values, colored lines represent actual values, grey shaded areas represent $95\%$ confidence intervals for the forecasts}%
\end{figure}

 From the plot, we can see wind energy slowly increasing, which is to be expected as California continues to build more wind farms. The only decreasing forecast is geothermal energy, which has been steadily declining since $2000$. Looking at the axes, we can again reemphasize the importance of solar and biomass in terms how much we consume compared to the other renewable sources. Biomass and solar are almost quadrupling the consumption of wind and geothermal energy. 

The next set of plots display the energy consumption in California by sector, each with their 10 year forecast. Forecasting consumption in the commercial sector utilized a (4,1,0) ARIMAX model that used California's population and the average price of energy in the commercial sector as exogenous predictors. Forecasting consumption in the residential sector utilized a (4,1,0) ARIMAX model that used California's population and the average price of energy in the residential sector as exogenous predictors. Forecasting consumption in industrial sector utilized a (2,1,2) ARIMAX model that used natural gas and petroleum prices as exogenous predictors. Forecasting consumption in transportation sector utilized a (2,1,2) ARIMAX model that used population and average energy price in that sector as exogenous predictors.

 \begin{figure}[H]%
    \centering
    \subfloat[\centering Transportation]{{\includegraphics[width=7.5cm]{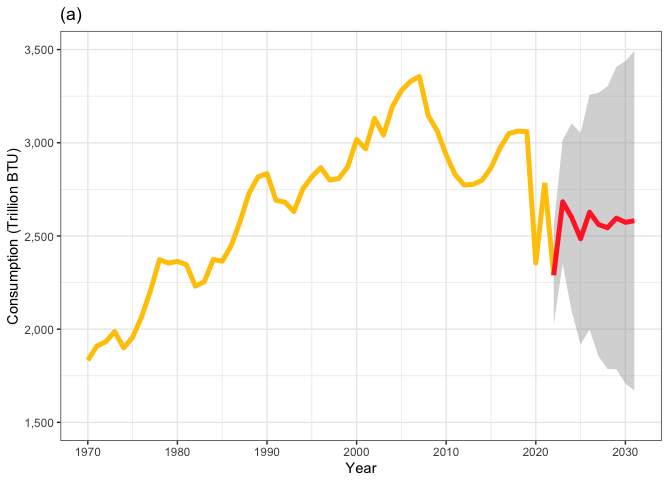} }}
    \qquad
    \subfloat[\centering Industrial]{{\includegraphics[width=7.5cm]{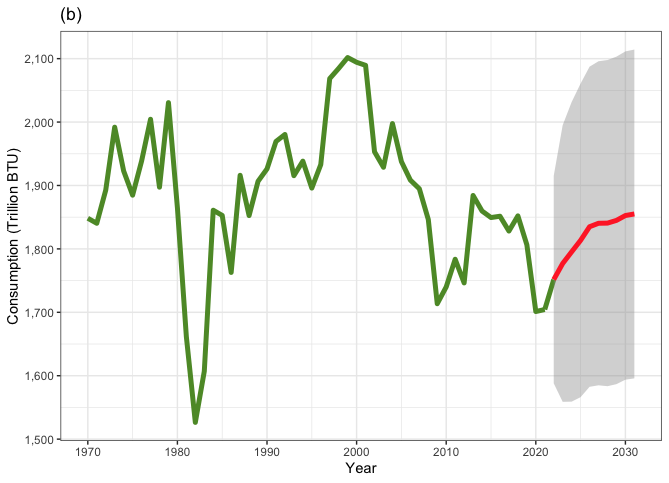} }}
    \newline
    \centering\subfloat[\centering Commercial]{{\includegraphics[width=7.5cm]{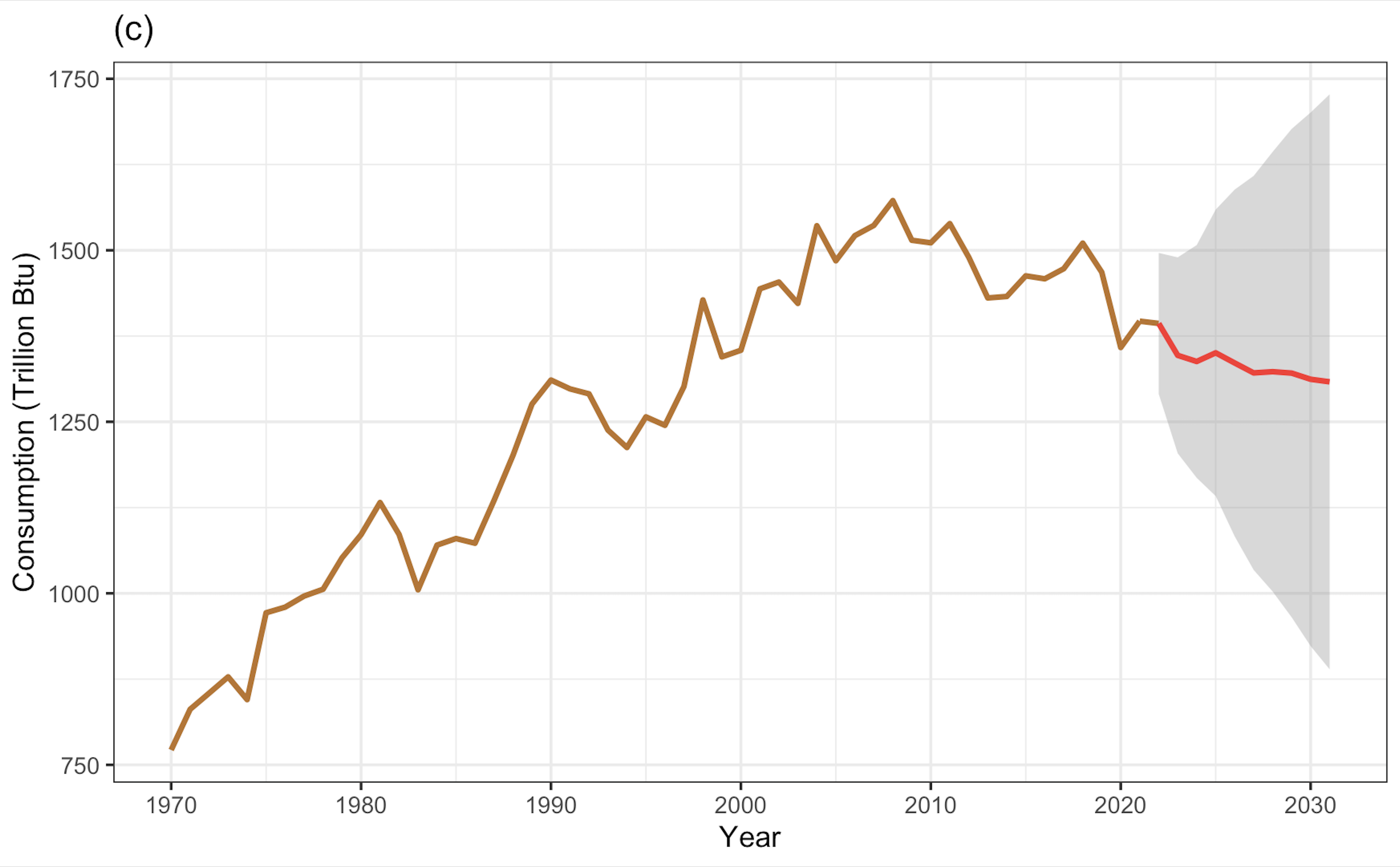}}}
    \qquad
    \subfloat[\centering Residential]{\includegraphics[width=7.5cm]{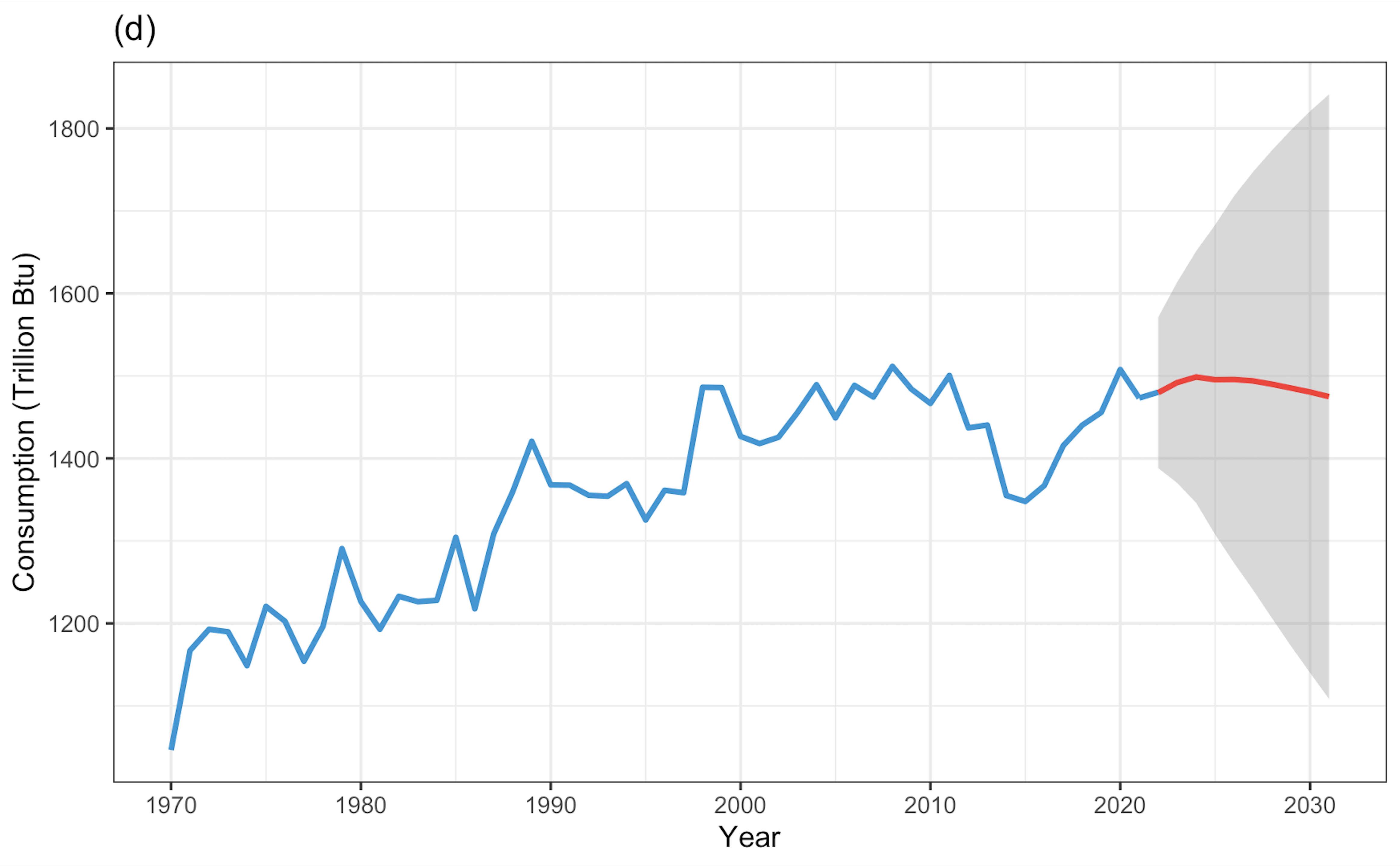}}
    \vspace{0.2cm}
    \caption{10 Year Energy Consumption Forecasts by Sector: (a) Transportation (b) Industrial (c) Commercial (d) Residential. Burgundy, blue, green, and yellow lines respectively represent the observed values for the previously listed sectors, red lines represent the forecast values, grey shaded areas represent 95\% confidence intervals for the forecast values}%
\end{figure}

From the above set of plots, we first predict the transportation sector to continue to recover from the drop induced by the COVID-19 pandemic and then begin to flatten out as time goes on. For the industrial sector, we predict consumption will increase, however, the rather large confidence band indicates anything can happen. We expect consumption in the commercial sector to slowly decline over the next decade. This can be supported by the decline in office usage and increase in the popularity of working from home following the start of the COVID-19 pandemic. Lastly, we expect residential energy consumption to remain somewhat stagnant as the population in California has begun to slightly decrease in recent years.

\newpage
\section{Discussion}

In all of our modeling approaches, we make the fundamental assumption of sustained economic stability throughout the entire forecast period and that no major natural catastrophes or wars will befall the state. However, it is important to acknowledge that the disruptive and unprecedented effects of the COVID-19 pandemic pose a significant challenge for ARIMA and ARIMAX models in effectively accounting for their impact on energy consumption. This challenge is further compounded by the limited availability of data for the year 2022, which was characterized by a return to near-normalcy.

Furthermore, certain exogenous variables in our analysis had fewer recorded years of data compared to the consumption variables. This limitation reduced the number of data points available for training our ARIMAX model. Despite these substantial challenges, our models exhibited commendable performance on the testing set, underscoring the accuracy and reliability of ARIMA and ARIMAX models in our forecasting framework.

\subsection{Future Work}
In the future, we would like to work with monthly data instead of annual data for California so we can also analyze the seasonality of energy usage across individual calendar years. Additionally, we would also implement a recurrent neural network to aid in our forecasting because research in this field shows that a neural network might account for the anomalies in the energy data better than an ARIMA or ARIMAX model.

\clearpage 
\printbibliography
\end{document}